\documentstyle[amssymb,aps,preprint]{revtex}

\tightenlines
\begin{document}
\title{{Mass generation for non-Abelian antisymmetric tensor fields in a
three-dimensional space-time}}

\author{
{\bf D. M. Medeiros$^{1,2}$, R. R. Landim$^{1}$, and C. A. S.
Almeida$^{1}$\thanks{E-mail address: carlos@fisica.ufc.br} }
\\
\vskip0.4cm
{\normalsize {\it $^{1}$Universidade Federal do
Cear\'{a} - Departamento de F\'{\i}sica }}
\\
{\normalsize {\it C.P. 6030, 60470-455 Fortaleza-Ce , Brazil }}
\\
\vskip0.4cm
{\normalsize {\it $^{2}$Universidade Estadual do Cear\'{a}
- Departamento de F\'{\i}sica e Qu\'{\i}mica }}
\\
{\normalsize {\it Av. Paranjana, 1700, 60740-000 Fortaleza-Ce,
Brazil }}}

\maketitle

\begin{abstract}
Starting from a recently proposed Abelian topological model in
(2+1) dimensions, which involve the Kalb-Ramond two form field, we
study a non-Abelian generalization of the model. An obstruction
for generalization is detected. However we show that the goal is
achieved if we introduce a vectorial auxiliary field.
Consequently, a model is proposed, exhibiting a non-Abelian
topological mass generation mechanism in $D=3$, that provides mass
for the Kalb-Ramond field. The covariant quantization of this
model requires ghosts for ghosts. Therefore in order to quantize
the theory we construct a complete set of BRST and anti-BRST
equations using the horizontality condition.
\end{abstract}
\vskip0.5cm
PACS\, 11.15.-q, 11.10.Kk,12.90.+b \vskip0.5cm

Antisymmetric tensor gauge fields provide a natural extension of
the usual vector gauge fields, appearing as mediator of string
interaction and having an important key role in supergravity.
Also, they are fundamental to the well known topological mass
generation mechanism \cite{Deser:1982wh} for Abelian vector boson
in four dimensions, through a BF term \cite{Allen:1991gb}. This
term is characterized by the presence of an antisymmetric gauge
field $B_{\mu \nu }$ ( Kalb-Ramond field) and the field strength
$F_{\mu \nu }.$ Non-Abelian extensions of models involving
antisymmetric gauge fields in four dimensional space-time were
introduced first by Lahiri \cite{Lahiri:1992hz,Lahiri:1997dm} and
later by Hwang and Lee \cite{Hwang:1997er}, in the context of
topologically mass generation models. Both procedures requires the
introduction of an auxiliary vector field, justified by the need
to untie the constraint between two and three form curvatures $F$
and $H$, and to the better of our knowledge, that is the first
approach in the literature to considering invariant non-Abelian
field strength for an antisymmetric tensor gauge field. A
non-Abelian theory involving an antisymmetric tensor field coupled
to a gauge field appears as an alternative mechanism for
generating vector bosons masses, similar to the theory of a heavy
Higgs particle. It is worth to mention a generalization to a
compact non-Abelian gauge group of an Abelian mechanism in the
context of non-Abelian quantum hair on black holes
\cite{Lahiri:1992yz}.

Kalb-Ramond fields arise naturally in string coupled to the area
element of the two-dimensional worldsheet \cite{Kalb:1974yc} and a
string Higgs mechanism was introduced by Rey in Ref.
\cite{Rey:1989ti}.

Recently, we have shown a topological mass generation in an
Abelian three-dimensional model involving a two form gauge field
$B_{\mu \nu }$ and a scalar field $\varphi $, rather than the
usual Maxwell-Chern-Simons model
\cite{Medeiros:1999df,Gomes:2001fr}. The action for the model just
mentioned reads as
\begin{equation}
S_{inv}^A=\int \!\!d^3x~\left( \frac 1{12}H_{\mu \nu \alpha }H^{\mu \nu
\alpha }+\frac 12\partial _\mu \varphi \partial ^\mu \varphi +\frac m2%
\epsilon ^{\mu \nu \alpha }B_{\mu \nu }\partial _\alpha \varphi \right) ,
\label{sa}
\end{equation}
where $H_{\mu \nu \alpha }$ is the totally antisymmetric tensor
$H_{\mu \nu \alpha }=\partial _\mu B_{\nu \alpha }+\partial
_\alpha B_{\mu \nu }+\partial _\nu B_{\alpha \mu }$.

The action (\ref{sa}), is invariant under the transformation
\begin{equation}
\delta \varphi =0,\quad \quad \delta B_{\mu \nu }=\partial _{[\mu }\omega
_{\nu ]}\text{ ,}  \label{invsa}
\end{equation}
and its equations of motion give the massive equations
\begin{equation}
(\square +m^2)\partial _\mu \varphi =0, ~ (\square +m^2)H_{\mu \nu
\alpha }=0 . \label{gmt061}
\end{equation}

The model described by action (\ref{sa}) can be consistently
obtained by dimensional reduction of a four-dimensional $B\wedge
F$ model if we discard the Chern-Simons-like terms
\cite{Medeiros:1999df}.

The purpose of this Brief Report is to construct a non-Abelian
version of the action (\ref{sa}). The only possibility is via an
introduction of an auxiliary vector field, as we have proved in
Ref. \cite{Medeiros:2001zm} using the method of consistent
deformations. We obtain here the BRST and anti-BRST equations by
applying the horizontality condition, including an auxiliary
vectorial field, which allows the sought non-Abelian
generalization. In this way, it is put in more rigorous grounds
the need to add an auxiliary field. In addition, we show a
non-Abelian topological mass generation mechanism for the
Kalb-Ramond field in three dimensions.

It is interesting to remark that the introduction of an one form
gauge connection $A$ is required to go further in the non-Abelian
generalization of our model (\ref{sa}), although our original
Abelian action (\ref{sa}) does not contain this field. Note that,
as pointed out by Thierry-Mieg and
Ne'eman \cite{Thierry-Mieg:1982nn} for the non-Abelian case, the field strength for $B$ is%
\footnote{%
Here and in the rest of the paper, in order to handle BRST
transformations, we use differential forms formalism for
convenience.}
\begin{equation}
H=dB+\left[ A,B\right] \equiv DB ,  \label{2.1}
\end{equation}
where $d=dx^\mu (\partial /\partial x^\mu )$ is the exterior derivative.

Resorting to Ref. \cite{Thierry-Mieg:1982nn}, we can define a new
${\cal H}$ given by
\begin{equation}
{\cal H}=dB+\left[ A,B\right] +\left[ F,C\right] , \label{2.2}
\end{equation}
where $C$ is the one form auxiliary field required and
$F=dA+A\wedge A$.

The obstruction to the non-Abelian generalization lies only on the
kinetic term for the antisymmetric field, but the topological term
must be conveniently redefined. So the non-Abelian version of the
action (\ref{sa}) can be written as
\begin{equation}
\int_{M_3}Tr\left\{ \frac 12{\cal H}\wedge ^{*}{\cal H}+m{\cal
H}\wedge \varphi +\frac 12D\varphi \wedge ^{*}D\varphi \right\} ,
\label{2.3}
\end{equation}
where $*$ is the Hodge star operator.

The action above is invariant under the following transformations:
\begin{equation}
\delta A=-D\theta , ~ \delta \varphi =\left[ \theta , ~ \varphi
\right] , ~ \delta B=D\Lambda +\left[ \theta ,B\right] , ~ \delta
C=\Lambda +\left[ \theta ,C\right] , \label{2.4}
\end{equation}
where $\theta $ and $\Lambda $ are zero and one form
transformation parameters respectively.

Here we shall use a formalism developed by Thierry-Mieg {\it et
al. }\cite {Thierry-Mieg:1982nn,Thierry-Mieg:1983un} in order to
obtain the BRST\ and anti-BRST transformation rules. In general
lines, we follow closely the treatment of Refs. \cite
{Thierry-Mieg:1982nn} or \cite{Hwang:1997er}, since the new object
introduced here, namely the scalar field, does not modify the
approach.

The presence of a scalar field in topological invariants is not so
uncommon. A three-dimensional Yang-Mills topological action was
proposed by Baulieu and Grossman \cite{Baulieu:1988xi} for
magnetic monopoles by gauge fixing the following topological
invariant:
\begin{equation}
S_{top}=\int_{M_3}Tr\left\{ F\wedge D\varphi \right\} .
\label{2.7}
\end{equation}

In the work of Thierry-Mieg and Ne'eman
\cite{Thierry-Mieg:1982nn}, a geometrical BRST quantization scheme
was developed where the base space is extended to a fiber bundle
space so that it contains unphysical (fiber-gauge orbit)
directions and physical (space-time) directions. Using a double
fiber bundle structure Quiros {\it et al. }\cite{Quiros:1981ec}
extended the principal fiber bundle formalism in order to include
anti-BRST symmetry. Basically the procedure consists in extending
the space-time to take into account a pair of scalar anticommuting
coordinates denoted by $y$ and $\overline{y}$ which correspond to
coordinates in the directions of the gauge group of the principal
fiber bundle. Then the so-called ''horizontality condition'' is
imposed. This condition enforces the curvature components
containing vertical (fiber) directions to vanish. So only the
horizontal components of physical curvature in the extended space
survive.

Let us define the following form fields in the extended space and
valued in the Lie algebra ${\cal G}$ of the gauge group:
\begin{equation}
\widetilde{\varphi }=\varphi ,  \label{2.8o}
\end{equation}
\begin{equation}
\widetilde{A}\equiv A_\mu dx^\mu +A_Ndy^N+A_{\overline{N}}d\overline{y}^{%
\overline{N}}\equiv A+\alpha +\overline{\alpha },  \label{2.8a}
\end{equation}
\begin{eqnarray}
\widetilde{B} &\equiv &\frac 12B_{\mu \nu }dx^\mu \wedge dx^\nu +B_{\mu
N}dx^\mu \wedge dy^N+B_{\mu \overline{N}}dx^\mu \wedge d\overline{y}^{%
\overline{N}}+\frac 12B_{MN}dy^M\wedge dy^N  \nonumber \\
&&+B_{M\overline{N}}dy^M\wedge d\overline{y}^{\overline{N}}+\frac 12B_{%
\overline{M}\overline{N}}d\overline{y}^{\overline{M}}\wedge d\overline{y}^{%
\overline{N}}  \nonumber \\
&\equiv &B-\beta -\overline{\beta }+\gamma +h+\overline{\gamma },
\label{2.8b}
\end{eqnarray}
and
\begin{equation}
\widetilde{C}\equiv C_\mu dx^\mu +C_Ndy^N+C_{\overline{N}}d\overline{y}^{%
\overline{N}}\equiv C+c+\overline{c} .  \label{2.8c}
\end{equation}

Note that we identify the components in unphysical directions with new
fields, namely, $\alpha $, $\beta $ and $c$ ($\overline{\alpha }$, $%
\overline{\beta }$ and $\overline{c}$) as anticommuting ghosts
(antighosts) and the commuting ghosts (antighost) $\gamma $ and
$h$ ( $\overline{\gamma }$ ). Since $B$ has 3 degrees of freedom
in 3 dimensions, the number of d.o.f. described by the set of
fields $B,$ $\beta ,$ $\overline{\beta },$ $\gamma ,$
$\overline{\gamma },$ and $h$ is 3-3-3+1+1+1=0. Obviously, the scalar field $%
\varphi $ has only one d.o.f.

The curvatures 2 form $\widetilde{F}$ and 3 form $\widetilde{{\cal
H}}$ in the fiber-bundle space are

\begin{equation}
\widetilde{F}\equiv \widetilde{d}\widetilde{A}+\widetilde{A}\wedge
\widetilde{A} , \label{2.9a}
\end{equation}
and
\begin{equation}
\widetilde{{\cal H}}\equiv \widetilde{d}\widetilde{B}+\left[ \widetilde{A},%
\widetilde{B}\right] +\left[ \widetilde{F},\widetilde{C}\right] ,
\label{2.9b}
\end{equation}
where $\widetilde{d}=d+s+\overline{s}.$ The exterior derivatives in the
gauge group directions are denoted by $s=dy^N(\partial /\partial y^N)$ and $%
\overline{s}=d\overline{y}^{\overline{N}}(\partial /\partial \overline{y}^{%
\overline{N}}).$

It is important to remark here that since we are focusing a mass generation
mechanism or, in other words, the action (\ref{2.3}), the extra symmetries
which appear in the pure topological model have no room in the present
discussion.

The horizontality condition, or equivalently, the Maurer-Cartan
equation for the field strength $F$ can be written as
\begin{equation}
\widetilde{F}\equiv \widetilde{d}\widetilde{A}+\widetilde{A}\wedge
\widetilde{A}=F ,  \label{2.10a}
\end{equation}
and for the 3 form ${\cal H}$ is
\begin{equation}
\widetilde{{\cal H}}\equiv \widetilde{d}\widetilde{B}+\left[ \widetilde{A},%
\widetilde{B}\right] +\left[ \widetilde{F},\widetilde{C}\right] ={\cal H}%
 .  \label{2.10b}
\end{equation}

Also we can impose the horizontality condition for the one form
$D\varphi $, which may be written as
\begin{equation}
\widetilde{D}\widetilde{\varphi }=\widetilde{d}\varphi +\left[ \widetilde{A}%
,\varphi \right] =D\varphi  .  \label{2.10c}
\end{equation}

By expanding both sides of Eq. (\ref{2.10a}) over the pairs of two
forms, one can obtain the following transformation rules:
\[
sA_\mu =D_\mu \alpha \text{ },\text{ }\overline{s}A_\mu =D_\mu \overline{%
\alpha } ,
\]
\begin{equation}
s\alpha =-\alpha \wedge \alpha \text{ },\text{
}\overline{s}\overline{\alpha }=-\overline{\alpha }\wedge
\overline{\alpha } ,  \label{2.11a}
\end{equation}
\[
s\overline{\alpha }+\overline{s}\alpha =-\alpha \wedge
\overline{\alpha } ~.
\]

In order to close the algebra, we introduce an extra scalar
commuting field $b$ valued in the Lie algebra ${\cal G}$ such that
\begin{equation}
s\overline{\alpha }=b ,  \label{2.11b}
\end{equation}
and consequently
\begin{equation}
\overline{s}\alpha =-b-\overline{\alpha }\wedge \alpha \text{ , }\overline{s}%
b=-\overline{\alpha }\wedge b , sb=0 .  \label{2.11c}
\end{equation}

On the other hand, expanding Eq. (\ref{2.10b}) over the basis of 3
forms yields
\[
sB_{\mu \nu }=-[\alpha ,B_{\mu \nu }]-D_{[\mu }\beta _{\nu ]}+[F_{\mu \nu
},c]\text{ },\text{ }\overline{s}B_{\mu \nu }=-[\overline{\alpha },B_{\mu
\nu }]-D_{[\mu }\overline{\beta }_{\nu ]}-[F_{\mu \nu },\overline{c}],
\]
\[
s\beta _\mu =-[\alpha ,\beta _\mu ]+D_\mu \gamma \text{ , }\overline{s}%
\overline{\beta }_\mu =-[\overline{\alpha },\overline{\beta }_\mu
]+D_\mu \overline{\gamma }~,
\]
\begin{equation}
s\overline{\beta }_\mu +\overline{s}\beta _\mu =-[\alpha ,\overline{\beta }%
_\mu ]-[\overline{\alpha },\beta _\mu ]+D_\mu h ~, \label{2.12}
\end{equation}
\[
s\gamma =-[\alpha ,\gamma ]\text{ },\text{ }\overline{s}\overline{\gamma }=-[%
\overline{\alpha },\overline{\gamma }]~,
\]
\[
\overline{s}\gamma +sh=-[\alpha ,h]-[\overline{\alpha },\gamma ]\text{ , }s%
\overline{\gamma }+\overline{s}h=-[\overline{\alpha },h]-[\alpha ,\overline{%
\gamma }] .
\]
Note that when we treat two odd forms, the $[$ $,$ $]$ must be reading as an
anticommutator.

The action of $s$ and $\overline{s}$ upon $c$, $\overline{c}$ and
$C$ is not defined in Eq.(\ref{2.12}). However, the condition
(\ref{2.10b}) leads us to
\begin{equation}
\widetilde{B}+\widetilde{D}\widetilde{C}=B+DC . \label{2.13}
\end{equation}

The condition (\ref{2.13}) yields the BRST and anti-BRST
transformations for the auxiliary field $C$ and its ghosts $c$ and
$\overline{c}$:
\[
sC_\mu =-[\alpha ,C_\mu ]+D_\mu c+\beta _\mu \text{ , }\overline{s}C_\mu =-[%
\overline{\alpha },C_\mu ]+D_\mu \overline{c}+\overline{\beta
}_\mu  ,
\]
\begin{equation}
sc=-[\alpha ,c]-\gamma \text{ ,
}\overline{s}\overline{c}=-[\overline{\alpha
},\overline{c}]-\overline{\gamma } ,  \label{2.14}
\end{equation}
\[
s\overline{c}+\overline{s}c=-[\overline{\alpha },c]-[\alpha ,\overline{c}]-h%
 .
\]

However, as usual, the action of $s$ and $\overline{s}$ on the ghosts and
antighosts is not completely specified by Eqs. (\ref{2.12}) and (\ref{2.14}%
). Therefore, a set of additional fields is required, namely, a
commuting
vector field $t_\mu ,$ two anticommuting scalar fields $\omega $ and $%
\overline{\omega }$ and a commuting scalar field $n$. These fields
are used to solve Eqs. (\ref{2.12}). Then, we get
\[
s\overline{\beta }_\mu =t_\mu  , \overline{s}\beta _\mu =-t_\mu
-[\alpha ,\overline{\beta }_\mu ]-[\overline{\alpha },\beta _\mu
]+D_\mu h,
\]
\[
sh=\omega  , \overline{s}\gamma =-\omega -[\alpha ,h]-[\overline{%
\alpha },\gamma ],
\]
\[
s\overline{\gamma }=\overline{\omega } , \overline{s}h=-\overline{%
\omega }-[\alpha ,\overline{\gamma }]-[\overline{\alpha },h],
\]
\begin{equation}
s\overline{c}=n , \overline{s}c=-n-[\alpha ,\overline{c}]-[\overline{%
\alpha },c]-h ,  \label{2.15}
\end{equation}
\[
st_\mu =s\omega =s\overline{\omega }=sn=0,
\]
\[
\overline{s}t_\mu =-[\overline{\alpha },t_\mu ]-[D_\mu \alpha ,\overline{%
\gamma }]-D_\mu \overline{\omega }-[\overline{\beta }_\mu ,t] , %
\overline{s}n=-[\overline{\alpha },n]-[\overline{c},b]+\overline{\omega }%
 ,
\]
\[
\overline{s}\omega =-[\overline{\alpha },\omega ]-[\alpha \alpha ,\overline{%
\gamma }]-[\alpha ,\overline{\omega }]-[h,b] , \overline{s}\overline{%
\omega }=-[\overline{\alpha },\overline{\omega }]-[\overline{\gamma },b]%
.
\]
The nilpotency of the $s$ and $\overline{s}$ operators was used to obtain
the last eight relations.

Finally, by expanding Eq. (\ref{2.10c}), we obtain
\begin{equation}
s\varphi =[\alpha ,\varphi ] , \overline{s}\varphi
=[\overline{\alpha },\varphi ] .  \label{2.16}
\end{equation}

Therefore, a complete set of BRST and anti-BRST equations, namely,
Eqs. (\ref {2.11a}-\ref{2.11c}), (\ref{2.14}-\ref{2.16}), and
(\ref{2.12}), associated with the classical symmetry defined by
Eq. (\ref{2.4}), was obtained.

It is important to point out the difference between the fields
which do not belong to the principal fiber
bundle expansion of the ''physical '' fields ($b,t_\mu ,n,\omega $ and $%
\overline{\omega }$) (introduced in order to complete the
BRST/anti-BRST algebra) and the auxiliary one form field $C$
introduced in order to overcome the obstruction to the non-Abelian
generalization. Note that here the {\it a priori }introduction of
the auxiliary field $C$, was necessary in order to fix the BRST
and anti-BRST transformation rules. Furthermore, the obstruction
to non-Abelian generalization of the four-dimensional BF model,
namely, the existence of
the constraint $[F,^{*}H]=0,$ appears in the context of our model as $%
[F,^{*}H-m\varphi ]=0$, as can be seen from the equations of motion of the
action (\ref{2.3}), considered in the absence of the auxiliary field.

The simplest scenario to study mass generation is to consider the
equations of motion of the action (\ref{2.3}). For convenience, we
define a new one form field as $K\equiv D\varphi$. Therefore, the
equations of motion can be written as
\begin{equation}
D^{*}{\cal H}=mK, ~ D^{*}K=-m{\cal H}.  \label{3.1}
\end{equation}
Equations (\ref{3.1}) can be combined into the following second
order equations:
\begin{equation}
\left( D^{*}D^{*}+m^2\right) {\cal H}=0, ~ \left(
D^{*}D^{*}+m^2\right) K=0 .
\end{equation}
Considering only linear terms for the fields, we get
\begin{equation}
\left( d^{*}d^{*}+m^2\right) H=0, ~ \left( d^{*}d^{*}+m^2\right)
d\varphi =0 , \label{3.5}
\end{equation}
which are similar to the Eqs. (\ref{gmt061}), and exhibit mass
generation for $H$ and $\varphi .$

On the other hand, by looking to the pole structures of the
propagators of the model, mass generation can also be established.
In order to obtain them, we use the action (\ref{2.3}) added with
convenient gauge fixing terms, namely
\begin{eqnarray}
S_T &=&\int_{M_3}Tr\left\{ \frac 12{\cal H}\wedge ^{*}{\cal H}+m{\cal H}%
\wedge \varphi +\frac 12D\varphi \wedge ^{*}D\varphi +\right.
\nonumber \\ &&\left. {\cal J}\text{ }\wedge ^{*}B+j\text{ }\wedge
^{*}\varphi +J \wedge ^{*}M+J_p\wedge ^{*}p+p\wedge ^{*}dM+M\wedge
^{*}dB\right\} , \label{3.7}
\end{eqnarray}
where ${\cal J}$, $J$ , $J_p$ and $j$ are currents related to the
fields $B,$ $M,$ $p$ and $\varphi $ respectively, which generate
propagators in the path integral formulation. The Lagrange
multiplier fields $M$ and $p$ are introduced in order to implement
the Landau gauge fixing.

Therefore, the tree-level effective propagators for the
Kalb-Ramond and scalar fields are
\begin{equation}
<\varphi \varphi >_{a,b}=-\frac{\delta _{ab}}{p^2-m^2} ~, \label{3.8}
\end{equation}
and
\begin{equation}
<BB>_{a\mu \nu ,b\rho \sigma }=\frac{\delta _{ab}}{p^2-m^2}\left[ g_{\mu
[\rho }g_{\sigma ]\nu }-\frac{g_{\mu [\rho }p_{\sigma ]}p_\nu }{p^2}+\frac{%
g_{\nu [\rho }p_{\sigma ]}p_\mu }{p^2}\right]~ ,  \label{3.9}
\end{equation}
where $a$ and $b$ are group indices, and $\mu ,\nu ,\rho $ and
$\sigma $ are space-time indices. It is interesting to note that,
here, the gauge field $B$ ''eats '' the scalar field (not a Higgs
field, however) and acquires a longitudinal degree of freedom and
a mass. The inverse process is possible too.

In this work we have succeeded in extending a tridimensional
Abelian topological model to the non-Abelian case. The model
considered here couples a second rank antisymmetric tensor field
and a scalar field in a topological way. We introduce two new
fields in the model in order to obtain the pursued non-Abelian
version. One field is a one form gauge connection ( $A$ ) which
allows us to define a Yang-Mills covariant derivative. The other
auxiliary field ($C$ ) is a vectorial one, which is required in
order to resolve the constraint that prevents the correct
nonabelianization.

A formal framework to consider the introduction of these fields
and the consequent new symmetries, is furnished by BRST and
anti-BRST transformation rules, which are obtained using the
horizontality condition. Although quite similar to other
topological models, it is worth to mention that, in this case, we
have constructed transformation rules for the Kalb-Ramond field,
for two one form fields and for a scalar field.

Here is worthwhile to mention that a similar mass generation model
was presented by Jackiw and Pi in Ref. \cite{Jackiw:1997jg}, where
a non-Abelian version of the mixed Chern-Simons term was
considered. However, a one form field was declared to carry odd
parity, so preserving the parity of the model. Besides, due to the
parity constraint, the Jackiw-Pi model has less gauge symmetries
than ours, and in these two features resides the essential
difference between the models considered. Furthermore, once Jackiw
and Pi change the two form field by one form field they have not
the obstruction to non-abelianization of the kinetic term detected
by Lahiri (four-dimensional case) and us (three-dimensional case).

Finally, the topological mass generation mechanism for an Abelian
model found out in a previous paper was extended for the
non-Abelian case, and we end up with an effective theory
describing massive Kalb-Ramond gauge fields in $D=3$ space-time.

We conclude mentioning the possible relevance of the present
discussion to string theory. Indeed, the Kalb-Ramond field couples
directly to the worldsheet of strings, and bosonic string
condensation into the vacuum realize the Higgs mechanism to the
Kalb-Ramond gauge field \cite{Rey:1989ti}. Therefore an
alternative scenario to give mass to the Kalb-Ramond field in the
context of strings may be an interesting continuation of our
present results.

This work was supported in part by Conselho Nacional de Desenvolvimento
Cient\'{\i }fico e Tecnol\'{o}gico-CNPq and Funda\c{c}\~{a}o Cearense de
Amparo \`{a} Pesquisa-FUNCAP.

\end{document}